# Analysing India's Cyber Warfare Readiness and Developing a Defence Strategy

Yohan Fernandes
Department of Computer Science and Engineering
Northumbria University
London, United Kingdom
yohan.fernandes@northumbria.ac.uk

Nasr Abosata
Department of Computer Science and Engineering
Northumbria University
London, United Kingdom
nasr.abosata@northumbria.ac.uk

*Abstract*— The demand for strong cyber defence measures grows, especially in countries such as India, where the rate of digitalization far exceeds cybersecurity developments. The increasing amount of cyber threats highlights the urgent need to strengthen cyber defences. The literature review reveals significant shortcomings in India's cyber defence readiness, especially in real-time threat detection and response capabilities. Through simulation models, the study explores network security behaviours and the impact of defences on network security. The next section of this study focuses on implementing a cyber threat detection system that uses machine learning to identify and categorise cyber threats in real time, followed by strategies to integrate it into India's present infrastructure. Also, the study proposes an educational framework for training cyber professionals. The study concludes with a reflection on the implemented defence strategies. It adds to the continuing discussion about national security by providing an in-depth investigation of cyber warfare preparation and recommending a systematic method to improving through both technological and educational solutions.

*Keywords—cyber defense, strategy, simulation, threat detection, education, India,*

## I. INTRODUCTION

In India, cyber security has seen massive improvements in terms of both research and applications. Despite the multitudes of research and the guidance of governing policies, India still continues to be one of the most targeted nations when it comes to cybercrime. This is due to the lack of security features in private and government organisations. As reported in the IBM Security Cost of Data Breach Report of 2022, India saw an average data breach cost of $2.2 million in 2022 [1]. Another contributing feature is the rapid digitalization of India, over the years access to internet has become simpler and more widespread. Thus the need for further research in the domain to improve cyber defense technology and the need for updated and future-proof strategies and governance policies is of critical importance.

The objectives of this study are to review international defense policies and strategies in the area of cyber warfare, understanding the current state of cyber defense structures around the world; to evaluate the feasibility and efficiency of India's current cyber defense policies and strategies with respect to emerging cyber threats; to investigate a comprehensive cyber warfare defense strategy for India by using simulation for research, automated real-time detection, and using gamification and simulation to implement training programs and determine the practicality of the proposed strategy; and to assess the proposed cyber defense strategy's possible impact on India's national security, and to recognize any possible issues or limitations that must be resolved.

The main contribution of this paper is to suggest a more resilient cyber defense strategy that can be adopted by India to defend itself from adversaries. This can be achieved by providing an improved defense strategy, more secure policies and a more effective way of educating and training individuals to protect themselves online. This can assist in safeguarding India's critical infrastructure, maintaining national and military security, and protecting the digital economy from downfall.

## II. LITERATURE REVIEW

The rapid development of cyber weapons highlights the potential for war and its effect on global stability. Nations now view cybersecurity as one of the most important security issues, and this recognition will only increase over time. There is no widely agreed definition of cyber warfare, the phrase is loosely used to describe events and activities in the cyber landscape.

Cyberwarfare as a domain is a part of extensive research and development, however, since it is a matter of national interest and thus has a lot of complexities involved. Robinson et al. [2] presented a definition model to help define what attacks can be considered to be cyber warfare attacks, which are different from cyber-attacks that have no military purpose based on their intent and the actor. The model proposed is somewhat effective in defining simple attacks based on algorithm set of if-else conditions. However, it is still not possible to determine the true intent of the event unless we know the motivation of the actor behind it. To correctly detect a cyberwarfare, attack the process implemented should be multidisciplinary, that includes not only technical but also legal, military, and political components. Goel [3] conducted a research study on attribution of attacks by using digital forensics, covering the need for attribution, advancements in the field, and challenges faced. Goel states that although digital forensic tools are easily available as open-source material and can be used to trace down actors, the anonymity of the internet makes it a difficult task as it involves factors such as faked or spoofed information. The downside of such attribution techniques is the lack of automation capability. Automation cannot be trusted because of false positives, and falsely declaring cyberwarfare might cause unnecessary tension.

When it comes to detection of attacks, Salim et al. [4] conducted a systematic literature review on different approaches used to detect Advanced Persistent Threats. Advanced Persistent Threats (APTs) are arguably the most popular cyber warfare technique used to gain access to networks for information, they operate discreetly and are unnoticed for an extended amount of time. Based on the review, Salim et al. presented an awareness model to detect and predict mobile APTs that follows the NIST Framework which works with Cyber-cognitive situation awareness (CCSA). Machine learning based intrusion detection system (IDS) is the most common method used for APT classification. However, these IDS are unable to detect APTs in real time because the behaviour of an APT changes continuously and need improved rules. Therefore, real time detection of cyber-attacks is essential to a good defense strategy.

The implementation of a framework and a set of policies is just the first step, the next and most important step is putting it into action. Ebert [5], conducted research that analyses the rise of cyber threats to India's national security over the last two decades and identifies the national measures implemented by the Indian government to combat these threats. The study reveals a growing gap between the creation and execution of cyber security legislation, which is due to political constraints, not wanting to engage in multistakeholder cooperation, and challenges in gaining benefits from global cyber security negotiations. Although India's cyber diplomacy and effort is advanced, the practicality and implementation of these efforts is limited. This is evidently true and makes India's cyber defense policy inefficient. This can be resolved by implementing a more decentralized approach that combines both the cyber team and the state may be necessary to effectively protect against cyber threats. To put simply, the overall cyber resiliency of the nation is a combined effort of the people, state and the technical team. The cyber defense of a nation should be built from the ground up so that the fundamentals are strong. In the case of cyberwarfare, adversary nations are less likely to target military and government infrastructure because of the prenotion that it is well secured. They are likely to make their way through common netizens and private sector organisations and systematically gain access to more critical information.

Sharma [6] provided a comprehensive analysis of India's approach to cyber warfare strategy in the next decade and how the defense should be strategic, operational and tactical. However, 8 years later, India's offensive cyberwarfare capabilities are still lacking when compared to China and other strongholds since they have taken a stance that is based on cyber deterrence. As stated by Sharma, it must develop cyber infrastructure and procedures simultaneously to accomplish national security objectives. This calls for education and training among citizens, which is fundamental. In agreement with Sharma, Kumar & Mukherjee [7] wrote a white paper published by Dion Global Solutions in 2013 that looks at cyber security in India from a skill development perspective. They stress that there is an urgent need for cyber security specialists in India. The paper puts forth suggestions such as developing cyber hubs and implementing a cyber literacy framework, but unfortunately to this date none of these are being implemented on a national scale. Based on the justifications mentioned in the paper we can conclude that a nationwide initiative to train personnel to work against cyber threats not only as part of a team but also individually is urgently required. Boopathi et al. [8]. He also mentions that here will be a demand for around 500,000 security experts over the next years. The strategy proposed will provide an improvised form of training and education to personnel.

The most recent cyber security regulation in India is The Digital Personal Data Protection Act 2023. It is not a strategy by definition, but it can be used to secure government and private organisations along with the general public, which are all susceptible to cyberwarfare attacks in one way or another [9]. Despite its achievements, India is still victim of multiple cyber-attacks. The Act faced criticism for possible strictness that might restrict innovation and for failing to adequately ensure individual privacy, especially given the discretionary powers it offers to the Central Government in the handling of personal data [10].

This justifies the need for a new cyber defense strategy in India that will incorporate means to bridge the gaps and meet all requirements for the present and near future. Firstly, it will provide a more accurate attack identification and attribution model and algorithm that works in real-time, which will work along with a response framework. Once the proposed cyber defense solutions and technology are in place, a gamification-based training and education framework will be implemented, which will be used to test the implementation of the cyber defense strategy that is investigated and to educate and train cyber security specialists.

### III. METHODOLOGY

The deliverable of this research is a cyber defense strategy for India that focuses on cyberwarfare defense. To make sure the strategy is effective, analysis of the proposed features is necessary. Thus, the research methodology will include scenarios-based simulations of cyber-attack events, that will be experimented and analysed for valuable information. This is followed by the training and development of an AI based threat detection model that works in real-time to identify threats. Lastly, an education and training framework will be implemented.

#### A. Simulation Models

Simulation based research can be helpful for resolving how or why questions regarding events outside the researcher's control [11]. The simulations will be carried out experimentally in a controlled testing environment using the software NetLogo, which is a multi-agent programmable modelling environment used for simulating scenarios based on variable factors. The first step is to design the test cases. The strategy proposed is a selection of informative cases, the idea of which is that if the conclusions are accurate for the given case, then it must also be valid for all other cases within the scenario [12].

#### B. Threat Detection Model

The threat detection model developed in this dissertation was designed using machine learning techniques. The model is trained on data from the CICIDS2017 dataset [13] which contains network capture data for a range of simulated attack scenarios. Using Google Cloud's Vertex AI and AutoML, the model optimises efficiency by automating the feature engineering and model selection processes. The model then needs to be implemented into the infrastructures in India's current defense.

#### C. Data Analysis

Data analysis of simulations involves a systematic evaluation of cases in a real-world environment in order to identify patterns, causes, and thus come to informed conclusions. This technique is especially effective in the field of cyber security, where knowing the outcomes of individual situations may give important theoretical and practical insights. Analysis is carried out based on the performance of the defense implementations in the set scenarios, which are based on real life situations.

*D. Legal and ethical Considerations*

The research method, implementations, test scenarios, and data usage must adhere to legal restrictions in the UK. The applicable legislations are The General Data Protection Regulation (GDPR), Data Protection Act 2018 (DPA 2018) and NIS regulations (The National Archives). The study must also comply with academic and research ethics. The collected data will only be used for this study and then deleted. Given the relation of the research to India, some Indian regulations such as the IT Rules (2021), The digital Personal Data Protection Act (2023) and the Personal Data Protection Bill (2019) are also applicable.

## IV. IMPLEMENTATION

*A. Simulation Models*

The models offer a simple but informative view of how elements, such as the number of attacks, the responsiveness of cyber defence measures, and the fundamental vulnerability of network nodes, interact within India's digital ecosystem.

This model simulates cyber-attacks intended to destabilize critical infrastructure networks. The model components as seen in Fig.1 are:

•Nodes (Circle): Critical infrastructure components that are part of the network. They have varying security levels.

•Threats (Bug): Attack agents that actively infect nodes. They are of three types: malware, phishing and DDoS, each having different nature.

•Control Center (House): Simulates the cyber security response operation. Mitigates attacks and recovers compromised systems.

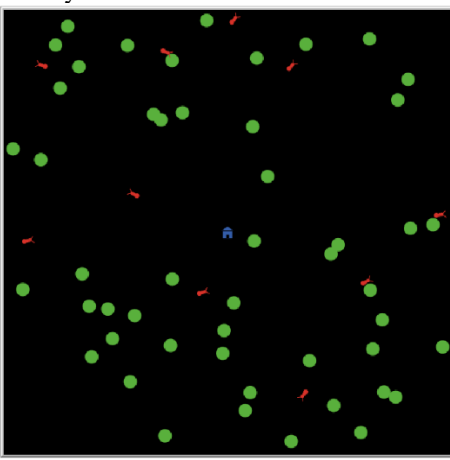

Fig. 1. Simulation Setup

*1) Scenario 1: Varying Number of Threats*

This run depicts a scenario in which the number of threats will be incremented from 10 to 30. Upon analysis this will show the capability of the network to handle a high number of threats and its scalability.

- Threat Levels: Random [1-3]
- Number of Threats: Varying [10, 30, 100]
- Defense Levels: Random [1-5]
- Control Centre Response Rate: 5
- Nodes: 50, Ticks: 200

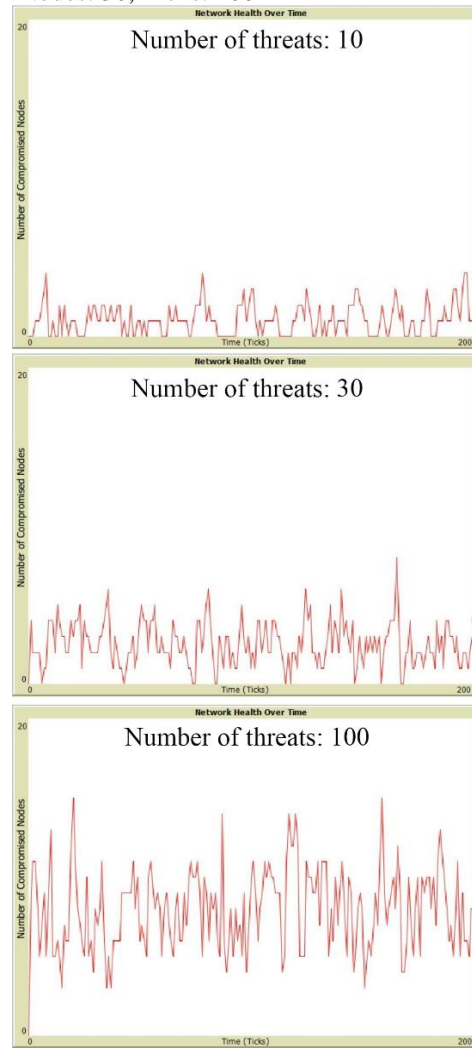

Fig. 2. Varying Number of Threats

As seen in fig.2, the network is able to handle 10 threats well, dealing effectively with ten concurrent attacks due to a standard response rate from the control centre. With 30 threats, the network's performance under load revealed an operational limit at which existing defences began to fail significantly. With 100 threats, which could be possible during targeted attacks or DDoS, the network was pushed to its limits revealing the need for improvement.

The tests suggest that as the number of attacks grows, the network may come across points of vulnerability where defence levels are lacking, particularly if the threats are concentrated or severe. The control centre's fixed reaction rate, when challenged by growing threats, emphasizes the importance of scalable and adaptable defence systems that can respond to changing situations without direct control centre intervention. The findings from this scenario can be used to guide network design improvements, identifying areas where extra resources will be needed to maintain security levels under high stress situations.

*2) Scenario 3: Varying Control Centre Response Rate*

This run depicts a scenario in which the control centre response rate will be increased from 2 to 10. Upon analysis this will show the impact of a strong control centre on the network. The objective is to see how changes in the control center's responsiveness affect overall network security and threat management.
- Threat Levels: Random [1-3]
- Number of Threats: 20
- Defense Levels: Random [1-5]
- Control Centre Response Rate: Varying [2, 8, 10]
- Nodes: 50, Ticks: 200

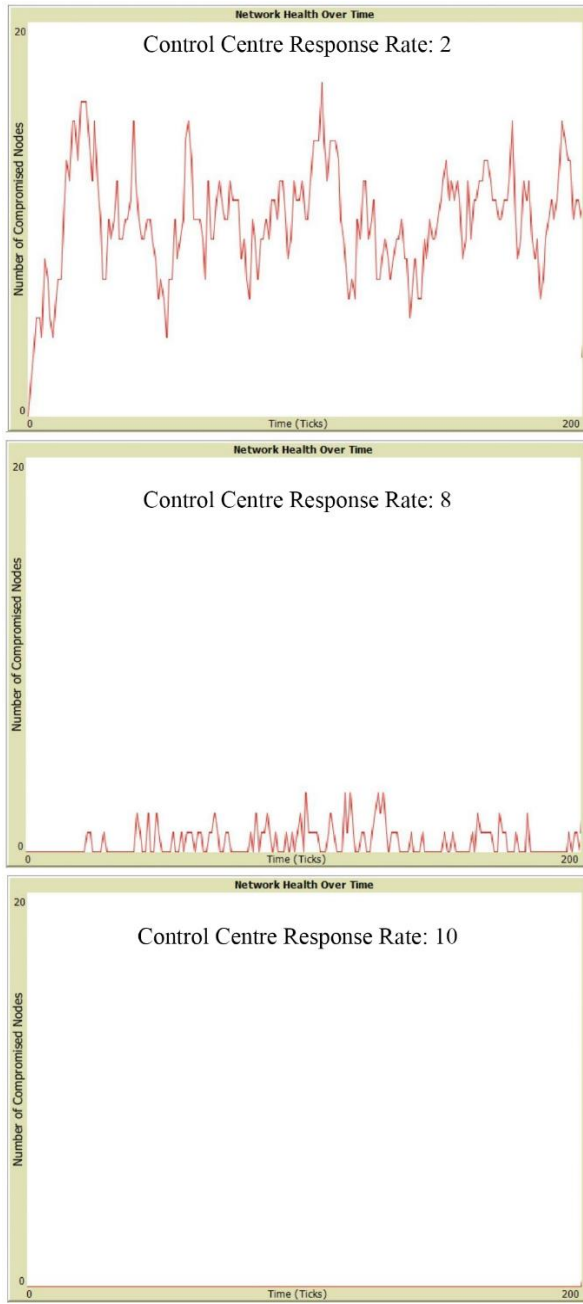

Fig. 3. Varying Control Centre Response Rate

As seen in Fig.3, a low control centre response rate of 2 showed very limited effectiveness in threat mitigation, indicating the need for faster response rate to maintain security. A response rate of 8 provided strong defence, efficiently coordinating defense throughout the network and significantly reducing the chance of intrusion. At the maximum response rate of ten, the network demonstrated optimum threat management while retaining high standards of functioning. No nodes were infected.

As the control centre's response rate increases, it's important to observe how quickly the network detects and mitigates threats. This suggests that a more proactive control centre could significantly improve network security. It is also important to identify any decreasing returns, where increases in reaction rate no longer result in improvements in threat management. This may indicate the point at which more investments in control centre responsiveness produce little returns and are hence not required.

*3) Scenario 3: Varying Defense Levels Among Nodes*

This run depicts a scenario in which the defense levels of the nodes will be increased from 1 to 5. Upon analysis this will show the impact of each node security on the network itself.
- Threat Levels: Random [1-3]
- Number of Threats: 30
- Defense Levels: Varying [1, 3, 5]
- Control Centre Response Rate: 3
- Nodes: 50, Ticks: 200

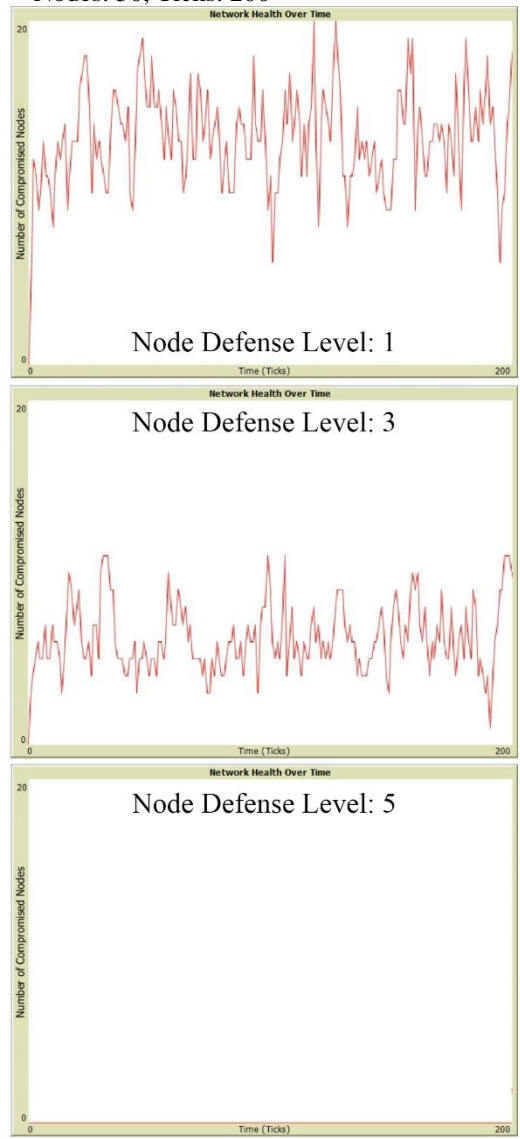

Fig. 4. Varying Node Defense Levels

With all nodes at the lowest defence level of one, the network was very vulnerable to attacks, stressing the need for a minimum feasible defence levels.

Defences at level 3 began to provide a balance, sufficiently guarding against the majority of threats, this is acceptable to an extent in low priority cases.

At the maximum defence level of 5, the network showed exceptional defensive capabilities, successfully defending against even the most severe attacks.

As predicted, stronger defence levels should correspond with increased threat resistance, resulting in fewer breaches and quicker threat management. The network's resilience as defence levels grow provides insight into the minimum defence necessary to preserve operational integrity under typical threat situations. The research also indicates that stronger node defences could reduce dependence on the control centre, as well as strategies for managing resources between node security and central response centre capabilities.

*4) Scenario 4: real-time Adaption Scenario*

This run depicts a real-time scenario in which the values are not set but are dynamically adjusted in real-time based on continuous threat assessment during the simulation. Upon analysis this will show the capability of the network to handle real-life situations and its ability to adjust to dynamic conditions. The simulation also has an added feature 'overall health' to accommodate real-time adaptation.

- Threat Levels: Random [1-3]
- Number of Threats: 10
- Defense Levels: Adapative (based on overall health of the network)
- Control Centre Response Rate: 5
- Nodes: 50, Ticks: 100

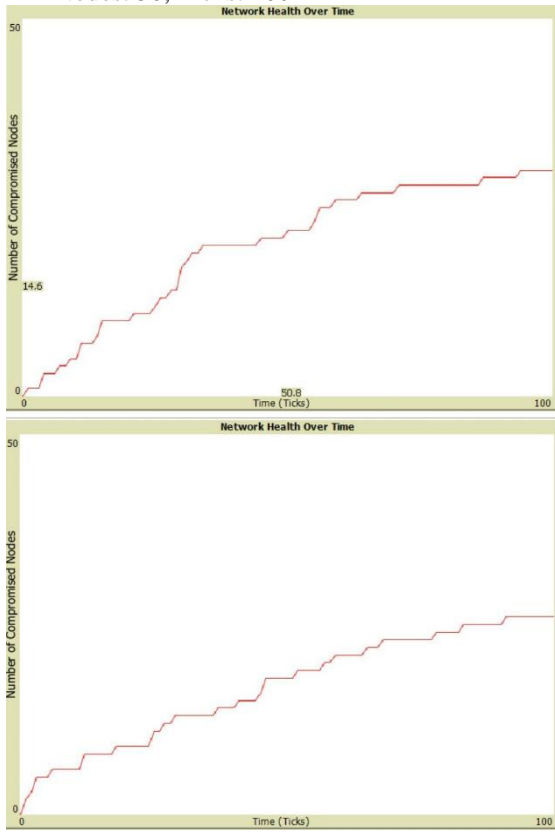

Fig. 5. Adaptive Defensive Levels

The ability to adjust defense levels In real-time essential for determining the network's capacity to automatically increase its defences in response to deteriorating conditions or to cut back when the risk level decreases. The efficiency of such adaptive systems may be evaluated by observing how the network responds to constant attacks with dynamic defences. It can be seen that networks that can adapt their defences in real time are more resilient to threats. The simulation highlights important thresholds at which the network's health affects its capacity to continue working properly. There will most certainly be an issue of balance between too frequent and insufficient modifications, since highly reactive systems may cause instability, while too inflexible systems may fail to adapt properly to increasing threats. Another reason for such a system is also the allocation of resources. Resources need to be divided among the control centre and each individual nodes, a self-adjusting system would help allocate resources efficiently.

*B. Cyber Threat Detection Model and Framework*

Generally, it is extremely difficult to recognize attackers and their objectives in cyberspace. This is because of the use of anonymity and obfuscation techniques. The solution must be automated so that there is no need for human interference, this will overcome the research gap mentioned by Goyal (2020). The algorithm and prediction model can be developed to resolve this issue as depicted in the figure.

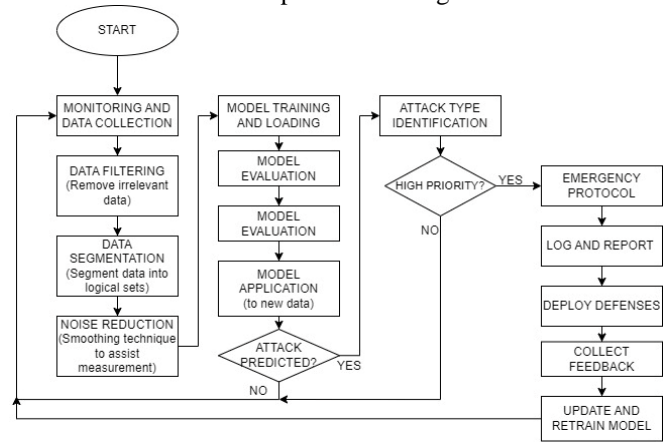

Fig. 6. Attack Detection and Identification Framework Model

The main objective of the model is to detect and categorize cyber risks based on data patterns and anomalies. It tries to improve cybersecurity by giving real-time alerts and practical knowledge that can help reduce the risks related to cyber-attacks. The input data used is the CICIDS2017 dataset is a popular cybersecurity resource that was put together mainly for developing and testing intrusion detection systems[27]. It includes a variety of simulated attack scenarios as well as regular traffic that was gathered over the course of one week. The collection contains popular attack types such as DDoS, port scanning, botnets and brute force. The data was cleaned and input into the training phase.

*1) Model Training*

Google Cloud Storage was used for storage. Vertex AI with AutoML was used for the development and deployment process. AutoML, or Automated Machine Learning, is a method or procedure for automating the stages in applying machine learning to real-world problems. AutoML tools perform different parts of data preparation automatically, including missing value handling, data normalisation, and encoding categorical variables. This lowers the requirement for manual intervention and data science knowledge. Classification modelling was used. AutoML examines multiple models and settings to determine which has the best performance. The model was trained using AutoML using the target column as 'Label' which denoted the type of the packet received i.e. if it is regular traffic or an attack.

*2) Model Evaluation*

Both the ROC AUC and PR AUC had values of 1.0. This means that the model can discriminate between distinct classes of vectors (regular or attack) with 100% accuracy under the test conditions. The model achieved a log loss of 0.001. Both the micro-average and macro-average F1 scores achieved a maximum of 1. This shows an exceptional balance between precision and recall. The following is the graphical evaluation analysis which depicts 100% precision with recall and maximum true positive rate with zero false positives.

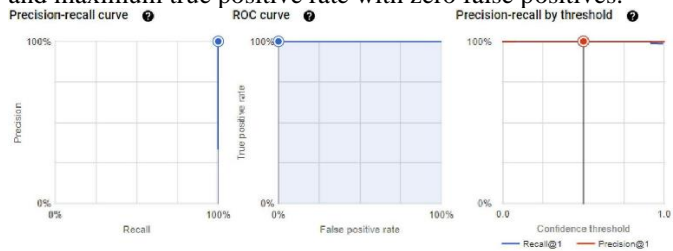

Fig. 7. Model Evaluation Charts

The model has performed in classifying the network traffic into three categories: BENIGN, FTP-Patator, and SSH-Patator. The model correctly identified all cases of benign traffic (43,166), FTP-Patator attacks (786), and SSH-Patator attacks (537). There were no cases where the model misidentified one type as another, suggesting excellent accuracy.

Fig. 8. Confusion Matrix

Understanding which features have the most important impact on the model's predictions is essential for refining the model and improving its interpretation and performance. The most influential features were found out to be:

- Source_IP: Most IP addresses are more likely to be associated with either benign or malicious traffic.
- Timestamp: The time at which network traffic occurs is essential, since it could indicate patterns or periods when attacks are more likely.
- Destination_IP: Similar to Source_IP, the destination IP address is also important for identifying potential targets in the network.
- Flow_Packets: The number of flow packets is essential, implying that the amount of data transmitted could indicate normal or suspicious behaviour.
- min_seg_size_forward: This feature indicates the minimum size of a segment in the forward direction, which may be indicative of the kind of traffic or protocol employed.

*3) Prediction*

The model is trained to predict whether the network traffic recorded is benign (regular network traffic) or malicious (an attack vector). The attack vector types are FTP-Patator or SSH-Patator, which are attack vectors originating from the penetration testing tool 'Patator' attacking the FTP server and SSH protocol respectively.

*4) Batch Prediction*

The input used for batch prediction is a data set collected from a different time period at the same experiment centre as the input data that the model was trained on.

Sample Prediction Outcome:

"prediction": {
"classes": ["BENIGN", "FTP-Patator", "SSH-Patator"],
"scores": [0.9557498693466187, 0.023342184722423553, 0.020907947793602943]}

Batch Prediction Details:

- Objective: Tabular
- Total Items: 288,602
- Predicted Items: 288,602
- Predictions: 288,602 succeeded, 0 failed
- Elapsed Time: 13min 35sec

*5) Model Integration and Framework*

The next step involves integrating the model created into the present cyber warfare security infrastructure present in India. This model aims to fill gaps within the infrastructure such as real-time detection and threat detection. The following framework will guide the implementation of the overall identification and defense process:

  i)   Data integration

The network traffic input to the model will be sourced from existing monitoring systems in India. The National Cyber Coordination Centre (NCCC) is an agency in India that is in charge of monitoring and managing communication data and intelligence gathering. When it comes to cyber warfare, the relevant input data can be obtained from agencies such as the Defence Cyber Agency (DCyA), which is tasked with cyber security threats in the military sector. This may need the development of APIs or infrastructure to allow for data exchange between systems such as NCCC's or DCyA's monitoring tools and the model. The detailed procedure involves deploying the model to an endpoint, which will provide an API that can be used for prediction. The secure API key is then retrieved and integrated into the monitoring system to set up external API calls. Code is implemented to send recorded data to the endpoint and receive predictions.

  ii)   Further Training and testing

The testing and deployment must be carried out in a controlled environment so that the active security is not effected. Scripts and available libraries can be used to automate the testing process and the resulting model needs to be reevaluated. In case of inaccuracy, data needs to be fine-tuned and additional data needs to be added if necessary. It is essential that the accuracy is high due to the nature of the use case of the model, especially in case of military purposes.

  iii)   Real-time alerts

The main principle of implementing real-time measures is make the entire process of identifying and reporting quicker. Threats have to be categorised based on the severity of their possible impact.

iv) SOP development

Standard Operating Procedures (SOPs) need to be developed in order to handle the identified threats. These are an essential part of the framework as they outline the plan and specific steps that need to be carried out for each kind of threat. The SOPs clearly define the roles and responsibilities of all stakeholders involved in cyber defence, this includes IT teams, security operations centres, and decision-makers inside organisations such as the DCyA and CERTs.

v) Feedback loop and training

The feedback loop is required to provide a systematic process for assessing the effectiveness of the prediction model and response strategies. The entire response process must be logged, this can be done by maintaining digital reports or filling in log forms. The reports and feedback is analysed and if required, improvements are discussed and implemented into the framework.

*C. Training and Education Framework*

The education and training modules offer a more effective way of preparing professionals working at all defence levels. Interaction and gamification elements are implemented into the training program to make it engaging and interesting. Presentations are a good way to teach, and quizzes are a good way to evaluate. This section will cover the use of a modified COFELET framework to evaluate the effectiveness of the educational modules for training cyber defence professionals. The COFELET framework is not specific to cyber security and hence requires adjustments.

*1) Framework Adaption*

The learning objectives of the framework are aligned with the needs of the matter i.e. the roles of cyber defense agents in the Indian Military and private sectors. This includes unique components such as cyber warfare attacks, threat analysis and practices involved in cyber security.

*2) Modifying Components*

The components of the COFELET framework are modified to adapt to the present research.

i) Scenario execution flows

The scenarios should be modified to include simulations and role-play scenarios of real-life cyber incidents. This would help learners apply their skills in a controlled but yet realistic environment, preparing them well for the real world.

ii) Challenge tuning

The challenge tuning component is used to modify complexity of the scenarios as per the learning level. In this case, the scenarios are modified by increasing stakes and creating situations that simulate high-stress environments. These levels are adjusted based on the level of the learner or the position of the learner within the team.

iii) Actions and tasks

The following tasks are to be implemented: red team-blue team exercises, forensic exercises, policy and regulation compliance training, leadership and management workshops (for management level learners), incident response drills, tabletop exercises.

*3) Evaluation*

The framework and its working is evaluated using the following processes.

i) Assessments

Assessments must be conducted both before and after training so that the effectiveness of the training is known. This can also be done by implementing gamification measures to measure the skill levels of learners in a quantitative manner.

ii) Feedback mechanism

Feedback from learners is important as it is key to helping with the refining process in order to improve the education modules. Feedback is collected in the form of questionnaires qualitatively.

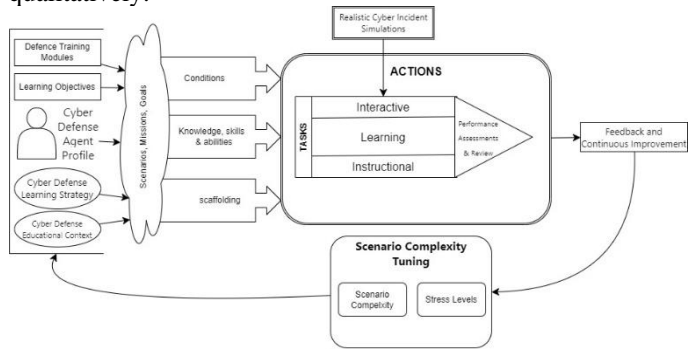

Fig. 9. Modified COFELET Framework

V. CONCLUSION

This research presents a comprehensive cyber defence approach that includes simulation for research, automated real-time detection using a detection model, and educational frameworks for improved personnel training.

The simulation scenarios in this work gave important insights into the dynamics of network security under a variety of scenarios. This study identified major weaknesses and strengths in networked systems by modifying key factors such as the number of threats, defence levels, and control centre reaction rates. The adaptive scenario, in particular, has shown the possibility of real-time, responsive security mechanisms that simulate real-world requirements. This research also contributes by developing a prediction model based on machine learning techniques. This technology is intended to identify cyber-attacks in real time when incorporated into the current monitoring and logging mechanisms in India's network infrastructure, considerably increasing the defensive capabilities of India's cyber defence system. The model's originality is its use of automated data analysis, which allows for a more nuanced understanding and response to possible cyber threats. Another notable addition is the development of an education and training framework to improve the practical abilities of cyber defence professionals. The framework created is a modification of the COFELET framework, which was altered to meet the specific requirements of the Indian cyber defence system. The modules use interactive situations to simulate real-world cyber challenges, offering learners with hands-on learning experiences. This method not only improves skill development, but it also prepares workers for the real-life cyber incidents.